\documentclass[10pt,letterpaper]{article}
\usepackage[top=0.85in,left=2.75in,footskip=0.75in]{geometry}

\usepackage{amsmath,amssymb}

\usepackage{changepage}

\usepackage[utf8x]{inputenc}

\usepackage{textcomp,marvosym}

\usepackage{cite}

\usepackage{nameref,hyperref}

\usepackage[right]{lineno}

\usepackage{microtype}
\DisableLigatures[f]{encoding = *, family = * }

\usepackage[table]{xcolor}

\usepackage{array}

\newcolumntype{+}{!{\vrule width 2pt}}

\newlength\savedwidth

\usepackage{tabularray}

\raggedright
\usepackage[aboveskip=1pt,labelfont=bf,labelsep=period,justification=raggedright,singlelinecheck=off]{caption}

\makeatletter
\renewcommand{\@biblabel}[1]{\quad#1.}
\makeatother

\usepackage{subcaption}

\usepackage{float}

\usepackage{pgfplots}
\pgfplotsset{compat=1.18}
\usepackage{graphicx}
\usepackage{dcolumn}
\usepackage{bm}
\usepackage{float}
\usepackage{tikz}
\usepackage{environ}
\makeatletter
\newsavebox{\measure@tikzpicture}
\NewEnviron{scaletikzpicturetowidth}[1]{%
  \def\tikz@width{#1}%
  \begin{lrbox}{\measure@tikzpicture}%
  \BODY
  \end{lrbox}%
  \pgfmathparse{#1/\wd\measure@tikzpicture}%
  \BODY
}
\makeatother
\usetikzlibrary{calc,angles,positioning,intersections}
\usetikzlibrary{decorations.pathreplacing,quotes}

\usepackage{lastpage,fancyhdr,graphicx}
\usepackage{epstopdf}
\fancyheadoffset[L]{2.25in}
\fancyfootoffset[L]{2.25in}
\newcommand{\figref}[1]{Fig.~\ref{#1} }

\usepackage[normalem]{ulem}

\graphicspath{{figs/}}

\begin{document}
\vspace*{0.2in}

\begin{flushleft}
{\Large
\textbf\newline{A Dynamic Model of Integration} 
}
\newline
\\
Joseph D. Johnson* \textsuperscript{1}
\\
Marisa C. Eisenberg \textsuperscript{2}
\\
\bigskip
\textbf{1} Department of Mathematics and Statistics, Carleton College, Northfield, Minnesota, United States
\textbf{2} Center for the Study of Complex Systems, University of Michigan, Ann Arbor, Michigan, United States
\\
\bigskip

%
%





* jjohnson16@carleton.edu

\end{flushleft}
\section*{Abstract}

Thomas Schelling introduced his agent-based model of segregation in 1971 and concluded that even when there is a low amount of intolerance within society that segregation will develop if people follow their individual preferences. A large body of literature building of this framework has been built and has bolstered this claim. This paper aims to take the same framework but instead look for ways to get to an integrated state. We focus on Allport's contact hypothesis that states that if there is equal status among groups, common goals among groups, and an institutional mechanism supporting intergroup contact then intergroup contact can reduce prejudice. We incorporate the contact hypothesis by having individuals adjust their intolerance based on their current neighborhood composition and the ease of conforming to their surroundings. Furthermore, we add in positive and negative media effects, as individuals are likely to get information about an outgroup from the media (e.g., news, TV, movies, etc.) that they consume. We find that having a society composed of individuals who do not easily conform to their surroundings and displaying positive examples of both groups in media promote integration within society.





\section*{Introduction}


Thomas Schelling created an agent-based model (ABM) of neighborhood segregation in 1971 \cite{schelling1971dynamic}. The model showed how individual preferences can lead to larger scale segregation within a population. Schelling's research revealed that segregation can occur even when individuals can tolerate being in the minority if agents move according to their individual preferences.

Many extensions to Schelling's segregation model have been implemented, adding components such as housing markets \cite{zhang2004dynamic,zhang2011tipping}, networks \cite{banos2012network}, reinforcement learning \cite{sert2020segregation} and external meeting places \cite{silver2021venues}. Many extensions of Schelling's work land on the idea that segregation is quite stable, with only a handful of papers exploring factors that lead to integration \cite{laurie2003role}.

Segregation has been proven to have negative effects on society\cite{kramer2009segregation}. For instance, it limits economic opportunities for Black individuals\cite{dickerson2007black}  and is associated with a reduction of access to healthy food options for Black individuals\cite{morland2002neighborhood,morland2007disparities}. Therefore, avoiding segregation should be a priority, and seeking potential solutions would benefit the public.

In this paper, we build off Schelling's segregation model in order to find situations where integration occurs. The main extensions included in this model are the effects of intergroup contact and media influence. Intergroup contact in this model is based on Allport's contact hypothesis \cite{allport1954nature}. Allport postulated a hypothesis that intergroup contact reduces prejudice when there is equal status among groups, common goals among groups, and an institutional mechanism supporting intergroup contact. In our model we will be considering contact by itself, assuming the other factors are already in place. 

This paper is not the first paper to incorporate adaptive tolerance and the contact hypothesis into their analysis of segregation. Urselmans and Phelps \cite{urselmans2018schelling} developed a model where agents became more tolerant if they were satisfied with the percentage of their ingroup and if that agent comes into contact at least one agent from their outgroup. Agents will decrease in tolerance if they are surrounded by agents from the outgroup. The amount the tolerance changes is fixed being equivalent for increases or decreases in tolerance and is set before the simulation begins.

The authors also included migration into their analysis, controlling the final proportion of migrants to native agents and the number of migrations waves (if the number of waves were greater than one then the amount of migrants each wave were adjusted to hit the desired migrant-to-native proportion).

Their work focused on migration and showed that under their framework that the tolerance will polarize to the extremes with agents becoming extremely tolerant or extremely intolerant. Here, the options for the equlibrium (ex: extremely tolerant, extremely intolerant or a mix of intolerant and tolerant) is determined by the amount that tolerance changes and proportion of native and migrant agents.

We believe our work adds to the body of work that analyzes segregation via agent-based modeling by varying the amount that agents adjust their tolerance based on that agent's neighborhood composition. We believe that this link between agents reaction and the neighborhood composition is plausible due to the survey results in the 1976 and 1992 Detroit Area Survey which find people's willingness to move to an area is dependent on the neighborhood current composition \cite{bruch2006neighborhood,farley1993continued}.


Frequently, individuals from different groups do not interact with one another directly. As a result, media has become a significant influence on people's perceptions of out-groups. Research has demonstrated that media can either increase or decrease an individual's prejudice levels, depending on how the out-group is portrayed\cite{das2009terrorism,shaver2017news} . To account for this, we have included mechanisms that can adjust an agent's tolerance level based on the type of media they consume.

Individuals move for various reasons outside of dissatisfaction with the racial or ethnic composition of their neighborhood. The Joint Center for Housing Studies (JCHS)  \cite{frost2020moving} states ``40 percent of movers did so for housing-related reasons in 2019, 27 percent moved for family-related reasons, 21 percent for job-related reasons, and 12 percent for other reasons''. Hence, the model also incorporates movement that is due to factors outside of the composition of an agent's neighborhood composition.

We recognize that the Schelling framework emphasizes individual mechanisms that cause segregation rather than societal factors that contribute to it, such as zoning laws, infrastructure that separates racial groups, and local racial covenants\cite{rothstein2017color,clark1969prejudice}. Our current investigation is introductory and lays the foundation for future work that builds upon this analysis.


\section*{Materials and Methods}


We create agents on a 51 $\times$ 51 grid. During the initial phase of the simulation, each grid cell has a probability, determined by the \textbf{density} parameter, of having one agent placed on it. The agent that is generated also has a probability of being either red or green, which is controlled by the \textbf{red-percent} parameter. The color of the agent indicates its ``racial'' group.


We take inspiration from the original Schelling model and assume agents have an intolerance level at timestep $t$, $I^t_j$, represented as the desired percentage of their neighbors who are from the same group. At the beginning of the simulation all agents are given the same intolerance is given by $I^0$. We define a neighborhood using the Moore neighborhood, which consists of the eight grid cells surrounding an agent. If the proportion of neighbors from the same group is lower than an agent's desired percentage, they will move to another location that meets their desired percentage. In our model, if no empty grid cells satisfy an agent's desired neighborhood composition, the agent will move to the empty space with the highest percentage of agents from the same group.


We have also incorporated a parameter, \textbf{non-racial-move}, which allows for a small chance of agents moving for reasons other than neighborhood composition, such as job changes, the housing market, or school systems. This addition reflects the reality that individuals do not exclusively base their moving decisions on neighborhood demographics. Alongside the probability of agents moving due to dissatisfaction with their neighborhood, there is an additional probability, at each time step, that agents will randomly move to an empty grid cell, disregarding neighborhood composition.

We borrow measures of segregation from Wilensky's NetLogo Segregation Model \cite{wilensky1997netlogo}. The level of segregation---assigned to variable \textbf{percent-similar}---is given by the mean number of agents that are similar to a given agent on a grid divided by average number of neighbors an agent has. This means that if one starts with a population split evenly between the two groups, an integrated equilibrium should have an average percent similar value close to 50 percent.

The stopping condition is determined by tracking \textbf{percent-similar} and the mean intolerance $\overline{I^t} = \sum_j I^t_j / N$---$N$ is the number of agents. We keep only the last 100 timesteps. If the standard deviation of \textbf{percent-similar} is within 1 and the standard deviation mean intolerance is within 0.5 then the simulation stops. There is a timescale $\tau$ called the adaptibility. 

Below, we lay out how intergroup contact and media effects will be implemented in the simulation.

\subsection*{Intergroup Contact}

As a starting point we borrow from Sabin-Miller and Abrams's work on political polarization \cite{sabin2020pull}. In their work, they develop an opinion model where individuals move towards political opinions that are close their own, but will be repulsed by opinions that are far away. We assume similar dynamics hold for intolerance values of the model's agents. In our model, agents compute the absolute difference of the intolerance $I^t_j$ to the current percentage of individuals in its Moore neighborhood that are from the same group $\%^t_j$ (this includes the agent itself). Specifically, the agents check if the absolute difference, $|\%^t_j-I^t_j|$, is bigger or smaller the ``conformity'' parameter $\lambda$.

Conformity, represented by the parameter $\lambda$, determines the degree of tolerance an agent has for the difference between its own intolerance and the neighborhood composition it desires. If the  absolute difference between the neighborhood percentage and its intolerance value is smaller than the conformity parameter, i.e., $|\%^t_j -I^t_j| < \lambda$ then the the intolerance value will: (1) increase if the neighborhood percentage is larger or (2) decrease if the neighborhood percentage is smaller.  If the  absolute difference between the neighborhood percentage and its intolerance value is larger than the conformity parameter, i.e., $|\%^t_j -I^t_j| > \lambda$, the intolerance value will: (1) increase if the neighborhood percentage is larger or (2) decrease if the neighborhood percentage is smaller. The way that the intolerance changes in response to neighborhood composition is outlined in Table \ref{tab:react}.

\begin{table}[H]
\centering
\begin{tabular}{|c|c|c|}
\hline
Sign of $\%^t_j-I^t_j$ & $|\%^t_j-I^t_j| < \lambda$ or $|\%^t_j-I^t_j| > \lambda$? & Sign of $f(\%^t_j,I^t_j)$ \\ \hline
Negative          &  Less Than                 & Negative             \\ \hline
Negative          & Greater Than                      & Positive             \\ \hline
Positive          & Less Than                        & Positive             \\ \hline
Positive          & Greater Than                      & Negative             \\ \hline
\end{tabular}
\caption{\textbf{How the reaction function $f(\%^t_j,I^t_j)$ works.} Non-boundary cases are illustrated in this table. If the absolute difference between the actual neighborhood composition and the desired neighborhood composition (intolerance) is smaller than the conformity then the change intolerance has the same sign.}
\label{tab:react}
\end{table}

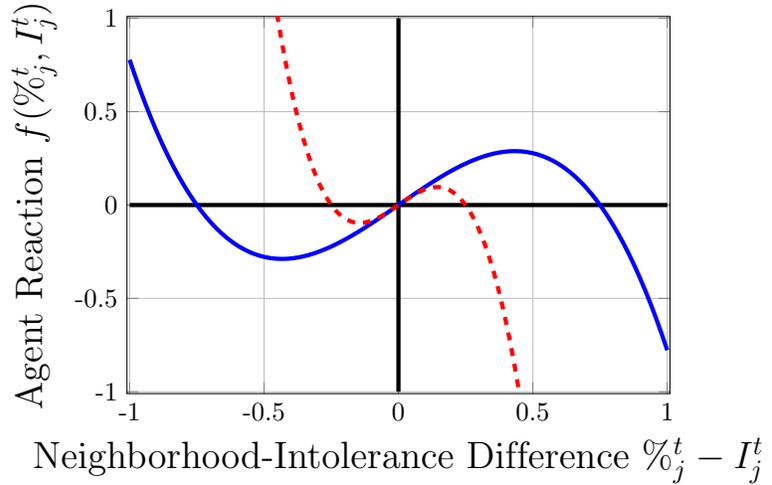
\begin{figure}[H]
    \centering 
    \begin{tikzpicture}
    \begin{axis}[width = 88mm, height = 66mm,grid=both,xmin =-1.01,ymin=-1.01, xmax =1.01, ymax =1.01, xtick={-1,-0.5,0,0.5,1}, xticklabels = {-1,-0.5,0,0.5,1},ytick={-1,-0.5,0,0.5,1}, yticklabels = {-1,-0.5,0,0.5,1},xlabel={{\Large Neighborhood-Intolerance Difference $\%^t_j-I^t_j$}},ylabel = {{\Large Agent Reaction $f(\%^t_j,I^t_j)$}}]
    \draw[ultra thick] (-1,0) -- (1,0);
    \draw[ultra thick] (0,-1) -- (0,1);
  \addplot[samples=100,blue,ultra thick, domain =-1:1]{ \x* (1 - (\x/0.75)*(\x/0.75) ) };
    \addplot[samples=100,red,ultra thick,  dashed, domain =-1:1]{ \x* (1 - (\x/0.25)*(\x/0.25) ) };
  \end{axis}
    \end{tikzpicture}

    \caption{\textbf{Sample reaction function.} Two cases of the reaction function, $f(\%^t_j,I^t_j)$\, with different conformity values: $\lambda = 0.25$ (red, dashed) and $\lambda = 0.75$ (blue, solid).}
    \label{fig:ExamplePlot}
\end{figure}

An agent's reactions to it's given surroundings will be given by the cubic function $f(\%^t_j,I^t_j)$, illustrated in \figref{fig:ExamplePlot}. I define $f(\%^t_j,I^t_j)$ as follows,
\begin{equation}
  f(\%^t_j,I^t_j) = (\%^t_j - I^t_j) \left(1 - \left(\frac{\%^t_j - I^t_j}{\lambda}\right)^2\right). \label{eq:Intol_React}
\end{equation}

We alter the reaction function so that intolerance values are kept within the range $I^t_j \in [0,1]$ by multiplying  $f(\%^t_j,I^t_j)$ by $I^t_j (1- I^t_j)$. This gives the final expression for the change in intolerance due to intergroup contact $\Delta I^t_j$, 

\begin{equation}
 \Delta I^t_j = I^t_j (1-I^t_j) f(\%^t_j,I^t_j)  = I^t_j (1-I^t_j) (\%^t_j - I^t_j)  \left(1 - \left(\frac{\%^t_j - I^t_j}{\lambda}\right)^2\right). \label{eq:Intol_Change}
\end{equation}

\begin{figure}[H]
    \centering 
    \begin{tikzpicture}
    \begin{axis}[width = 88mm, height = 66mm,grid=both,xmin =0,ymin=-0.101, xmax =1.01, ymax =0.101, xtick={-1,-0.5,0,0.5,1}, xticklabels = {-1,-0.5,0,0.5,1},ytick={-0.1,-0.05,0,0.05,0.1}, yticklabels = {-0.1,-0.05,0,0.05,0.1},xlabel={{\Large Intolerance $I^t_j$}},ylabel = {{\Large Intolerance Change $\Delta I^t_j$}}]
    \draw[ultra thick] (-1,0) -- (1,0);
    \draw[ultra thick] (0,-1) -- (0,1);
  \addplot[samples=100,blue,ultra thick, domain =0:1]{ \x*(1-\x)*(\x- 0.5)* (1 - ((\x- 0.5)/0.75)*((\x- 0.5)/0.75) ) };
    \addplot[samples=100,red,dashed,ultra thick, domain =0:1]{ \x*(1-\x)*(\x- 0.5)* (1 - ((\x- 0.5)/0.25)*((\x- 0.5)/0.25) ) };
  \end{axis}
    \end{tikzpicture}

    \caption{\textbf{Example change function.} The change in the tolerance value $\Delta I^t_j$ given by Eq. is plotted here. Here, half of the $i^{th}$ agent's neighborhood is composed of individuals from its own group, i.e., $\%^t_j = 0.5$. The conformity values are set to $\lambda = 0.25$ (red, dashed) and $\lambda = 0.75$ (blue, solid).}
    \label{fig:ExampleChangePlot}
\end{figure}
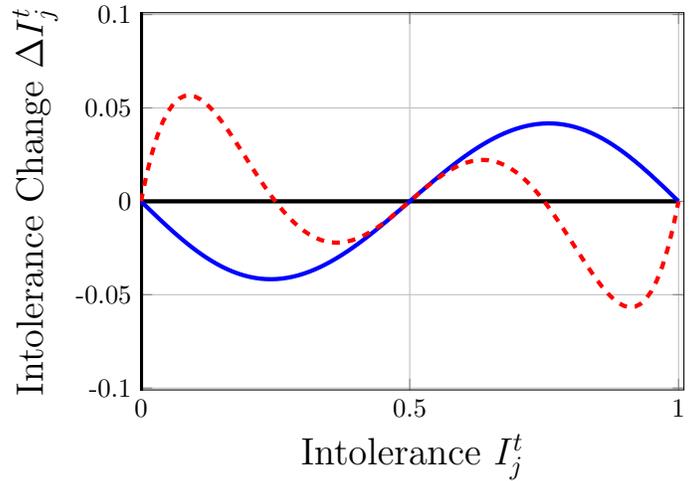

Since intolerance values are bounded within the range $I^t_j \in [0,1]$, the magnitude of the differences between intolerance and neighborhood composition cannot be larger than 1 and hence the conformity values are bounded within the range $ 0 < \lambda < 1$ through each simulation. We present two examples of $\Delta I^t_j$ in \figref{fig:ExampleChangePlot} when $\lambda = 0.75$ (blue, solid) and $\lambda = 0.25$ (red, dashed) assuming that 50 percent of the $i^{th}$ agent's neighborhood is of the same type as agent $j$ (i.e, $\%^t_j = 0.5$).

\begin{figure}[H]
    \centering
    \includegraphics[width=\textwidth]{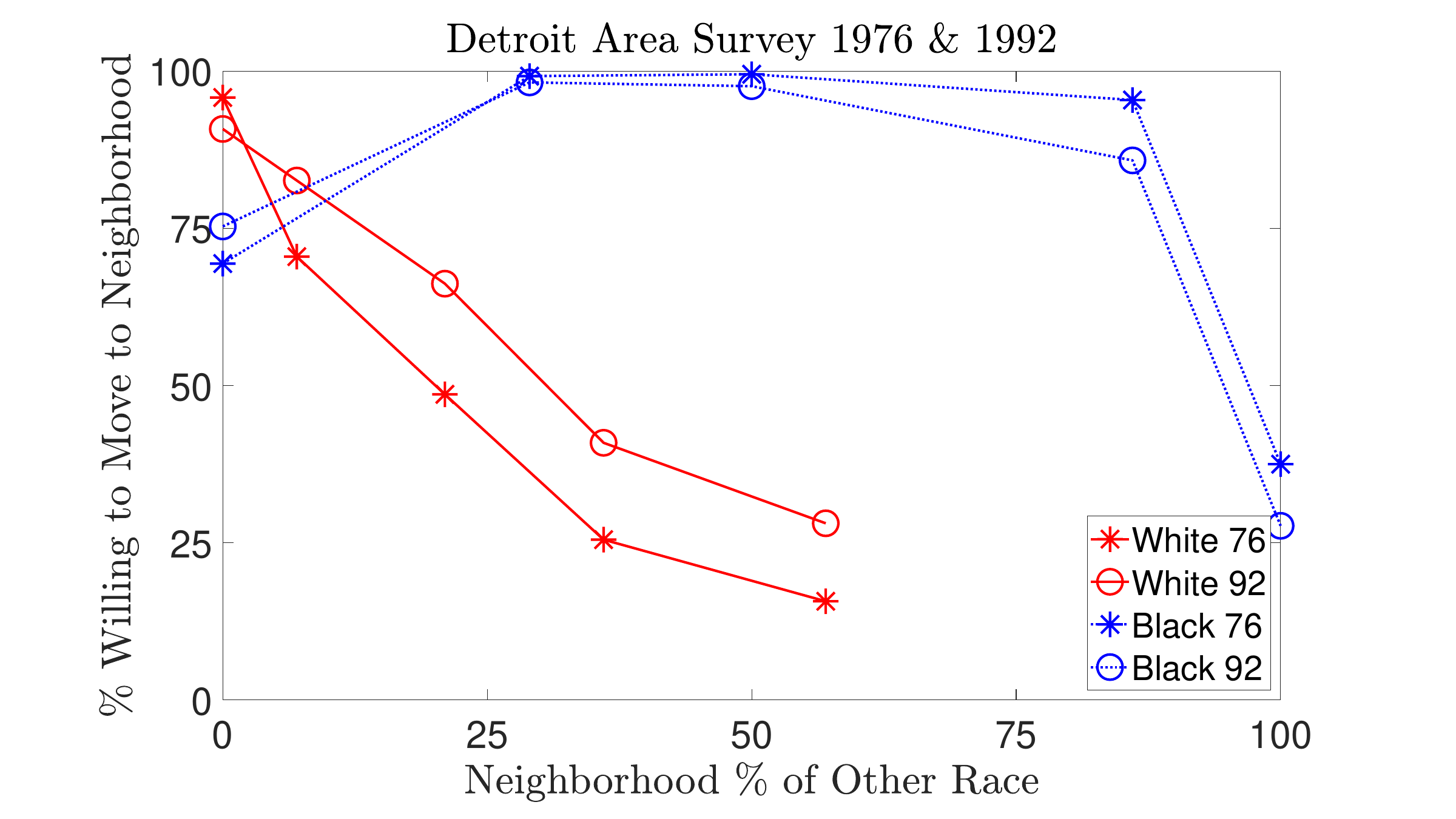}
    \caption{\textbf{Results from the Detroit Area Survey.} We plot the response data from the 1976 and 1992 Detroit Area Survey. Participants in this survey were asked if they were willing to move to a prospective house given the percentage of the ``other'' race---Black for White individuals and White for Black individuals---in the neighborhood.  We can see that the percentage of White respondents quickly drops as the percentage of the other race rises. The responses from Black individuals were U-shaped, where a neighorhood with a 50-50 split between White and Black people were the most preferred and all-Black neighborhoods or all-White neighborhoods were not as desired.}
    \label{fig:DAS}
\end{figure}

It is unlikely that we have formulated the exact function for how individuals adjust their tolerances. However, we aim to explore various other functions in future work. Nonetheless, if we assume that the Contact Hypothesis holds some validity, our reaction function might align with the results given in the Detroit Area Study and research conducted by Farley et al. \cite{bruch2006neighborhood,farley1993continued}. The survey results are portrayed in \figref{fig:DAS}. These studies suggest that Black individuals are open to living in integrated areas and may even prefer integration over a purely Black neighborhood. In contrast, White individuals tend to live in almost purely White neighborhoods and have less contact with Black individuals. As a result, integration may be less appealing to them. We conjecture that this may be due to the fact that Black individuals come into contact with White individuals more frequently compared to White people.

\subsection*{Media Effect}

Media bias towards certain groups of people is a common phenomenon. For example, Black and Latin individuals in America are overly represented as criminals in media and White individuals are overly represented as upholders of the law \cite{dixon2000overrepresentation}. We implement the effect of this bias on the $i^{th}$ agent's intolerance as a multiplicative factor, $g(j)$, in the following fashion
 \begin{equation}
    I^{t+1}_{j} = (1-g(j)) (I^t_j + \Delta I^t_j).
\end{equation}

If $ 0 < g(j) \leq 1 $, the agent is consuming \textbf{positive (tolerant)} media about the outgroup and hence their intolerance is decreasing. If $g(j) < 0$, the agent is consuming \textbf{negative (intolerant)} media about the outgroup and hence their intolerance is increasing.  Finally, if $g(j) = 0$, the media is \textbf{neutral} and thus their intolerance does not change. It is important to note also that the media effect modulates the impact of an individuals local neighborhood change as well (as $g(j)$ is implemented as a multiplicative rather than additive effect).



\begin{table}[H]
\centering
\resizebox{\textwidth}{!}{\begin{tabular}{cc}
    \hline
    Parameter   & Description \\ 
                \hline \hline
    density         &  The expected proportion of the grid that will be occupied by an agent \\ 
                \hline
   red-percent    & The percentage of the population from the red group \\ 
                    \hline 
    non-racial-move    & The probability that an agent will move to a random empty square at each time step. Ignores neighborhood composition. \\ 
                    \hline
    $I^t_j$     &  The minimum proportion from the same group in an agent's neighborhood that satisfies that agent\\ 
                        \hline
    $I^0$     &  Initial minimum proportion from the same group in an agent's neighborhood that satisfies that agent \\ 
                \hline
    $\%^t_j$       & The proportion from the same group in an agent's neighborhood \\
                \hline
    $\lambda$  &  The willingness for an agent to conform to its neighborhood      \\
                    \hline
    $g_(j)$  &  Media consumed by agent $j$   \\
    \hline
\end{tabular}}
\caption{\textbf{Parameter definitions.} Table of parameters used in the model with descriptions.}
\label{tab:param}
\end{table}

\section*{Results}



\subsection*{Conformity}

We investigate how the agents' conformity affects equilibrium values for segregation.  In the following simulations, we set the split between the groups to be 50-50, the \textbf{non-racial-move} probability to 5 percent and the \textbf{density} to 87.5 percent.

\begin{figure}[H]
    \centering
    \includegraphics[width = \textwidth, height = 75mm]{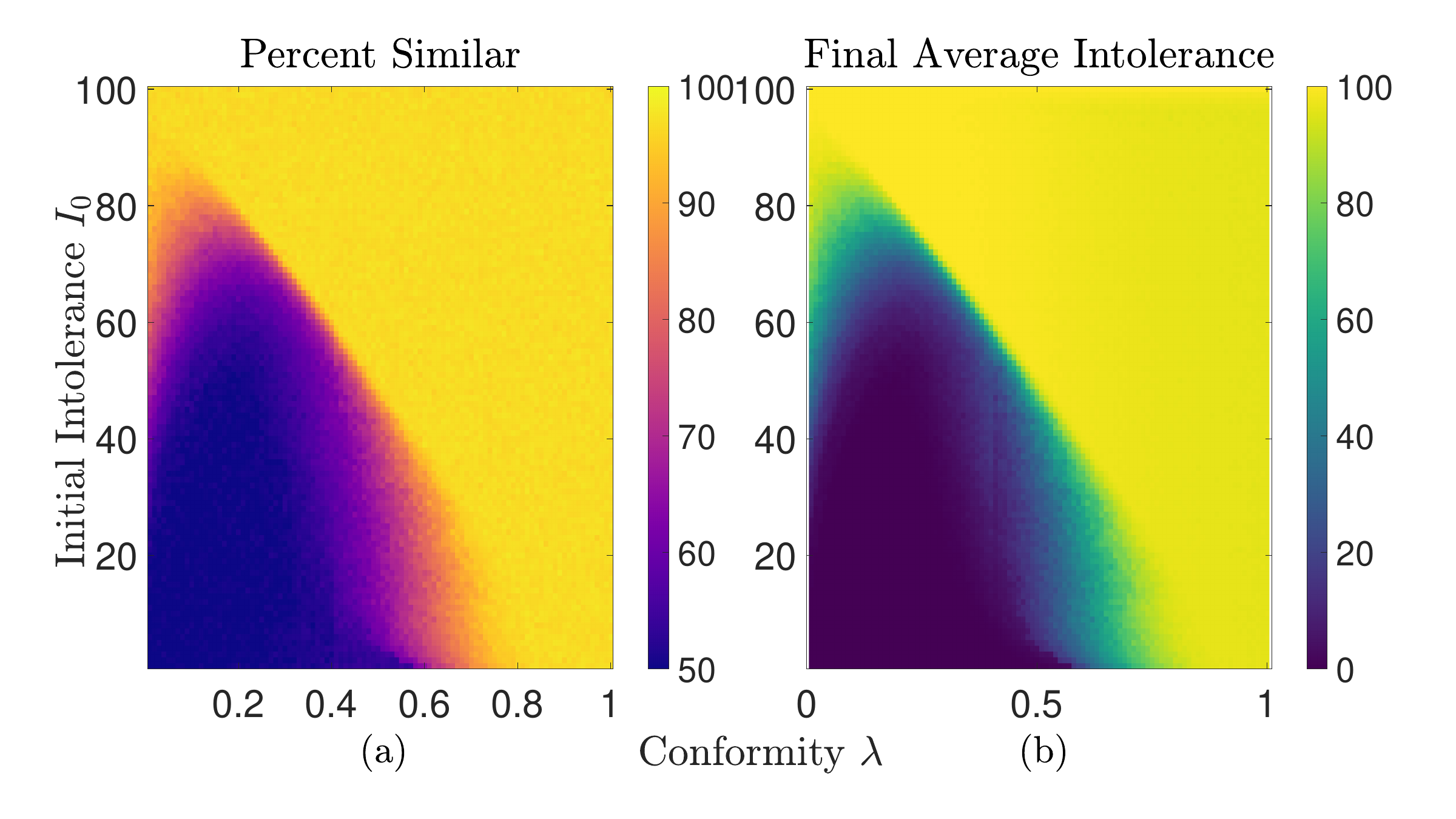}
  \caption{\textbf{Heatmap of Segregation and Average Final Intolerance.} (a) Equilibrium values of segregation measured by percent-similar, and (b) final average intolerance as the conformity $\lambda$ and initial intolerance $I^0$ varies. For all values of $I^0$, fully segregated equilibria occur for past some threshold $\lambda^*(I^0)$. Integrated equlibria occur when conformity values below $\lambda^*(I^0)$. Here, \textbf{red-percent} = 50, \textbf{density} = 0.875, and \textbf{non-racial-move} = 0.05. }
    \label{fig:PercentSimandFinIntol}
\end{figure}

The effect of conformity $\lambda$ on segregation is given in panel (a) of \figref{fig:PercentSimandFinIntol}. For any initial value of intolerance small values of conformity lead to integrated equilibria. Integration persists until a critical level of conformity, $\lambda^*(I^0)$, that depends on the initial intolerance. This threshold appears to be nonlinear in nature when starting in a random configuration.



We performed the same experiment with a completely segregated initial condition. This segregated initial condition was generated by spawning all red agents at the bottom of grid and then spawning green agents above the red agents.  This threshold becomes linear, $\lambda^* = 0.99-I^0$ when the simulation starts in completely segregated state. 

The reason why the relationship is linear is because agents initial neighborhood configuration is $\%^0_j = 1$ for all $j$ aside from the ones at the boundary of the red and green agents for both kinds of agents. This means that if $|1 - I^0| > \lambda$ then the most agents will have an adverse reaction to the segregated condition and their intolerance will decrease. Since there are only random swaps at the beginning many agents will spend time in relatively segregated neighborhoods, leading to a further decrease in intolerance. Then, the random movement mixes the population resulting in an integrated equilibrium.

\begin{figure}[H]
    \centering
    \includegraphics[width = \textwidth, height = 75mm]{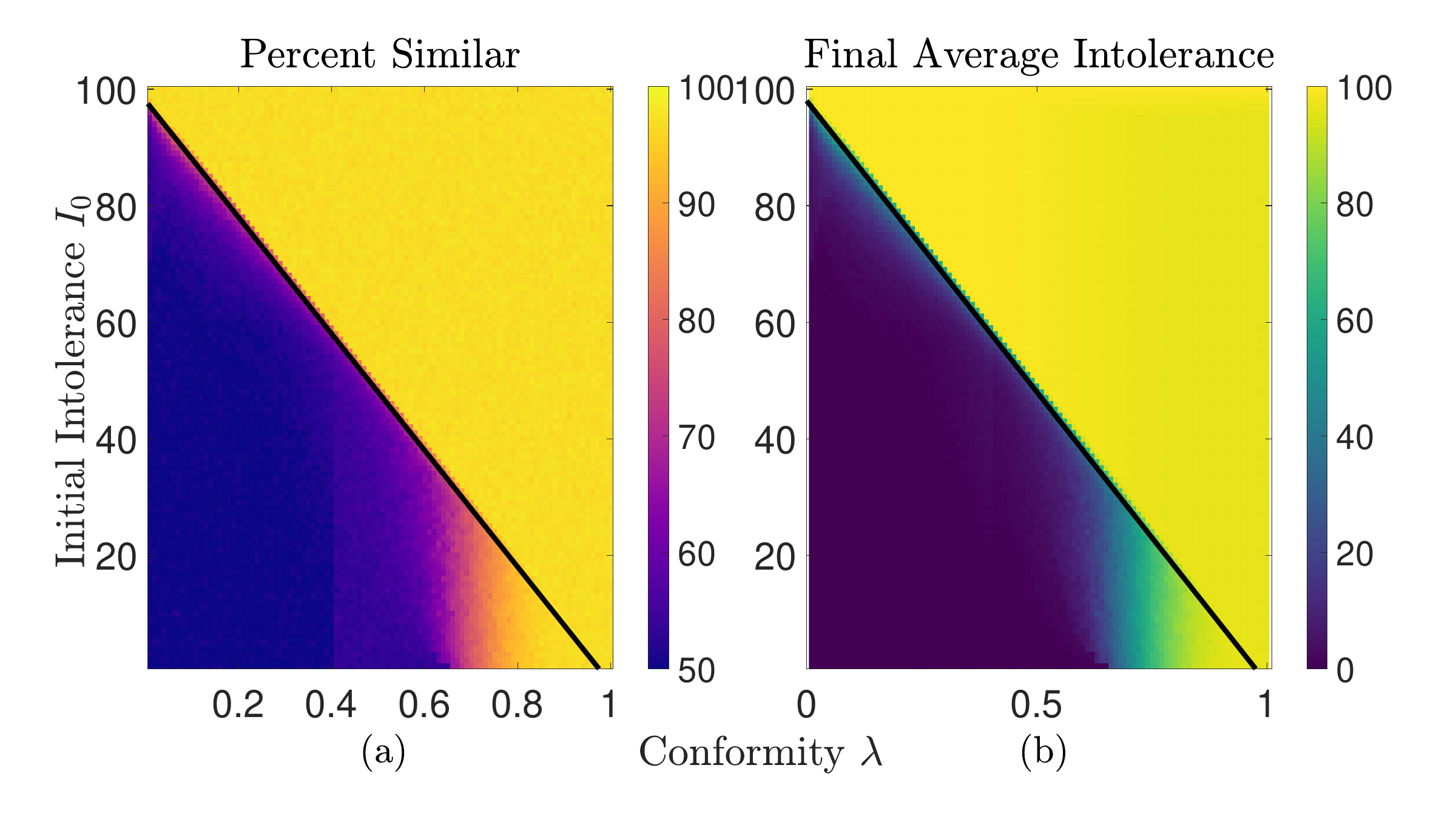}
   \caption{\textbf{Heatmap of Segregation and Average Final Intolerance for Segregated Initial Condition.} (a) Equilibrium values of segregation measured by percent-similar, and (b) final average intolerance as the conformity $\lambda$ and initial intolerance $I^0$ varies. Agents start in a segregated state, silo-ed in neighborhoods of all red agents or all green agents (except for agents at the boundary). For all values of $I^0$, fully segregated equilibria occur for past some threshold $\lambda^*(I^0) = 0.99 - I^0$ (black line). Integrated equlibria occur when conformity values below $\lambda^*(I^0)$. Here, \textbf{red-percent} = 50, \textbf{density} = 0.875, and \textbf{non-racial-move} = 0.05. }
    \label{fig:PercentSim_Compare}
\end{figure}

The final average intolerance of the agents match the segregation results. When $\lambda < \lambda(I^0)$ the intolerance drops near zero and will accept neighorhood composition, matching the integrated state. When $\lambda$ is above this threshold the agents desire neighborhoods composed purely of their own kind.  Therefore, segregation develops when individuals largely conform to their surroundings and integration develops when individuals do not readily conform to their surroundings.



\subsection*{Population Ratio}
We next investigated whether the results of the previous section are robust to differences in the population ratio of agent types. 

In real world settings, populations are often not equally sized, but rather there may be differing population sizes that can affect the patterns of segregation observed \cite{galster1990white,kye2018persistence}. We evaluated whether the results of the previous subsection are robust as the population ratio varies. In the following simulations we set the initial intolerance to $I^0 = 25, 50$.

\begin{figure}[H]
    \centering
    \includegraphics[width = \textwidth, height = 75mm]{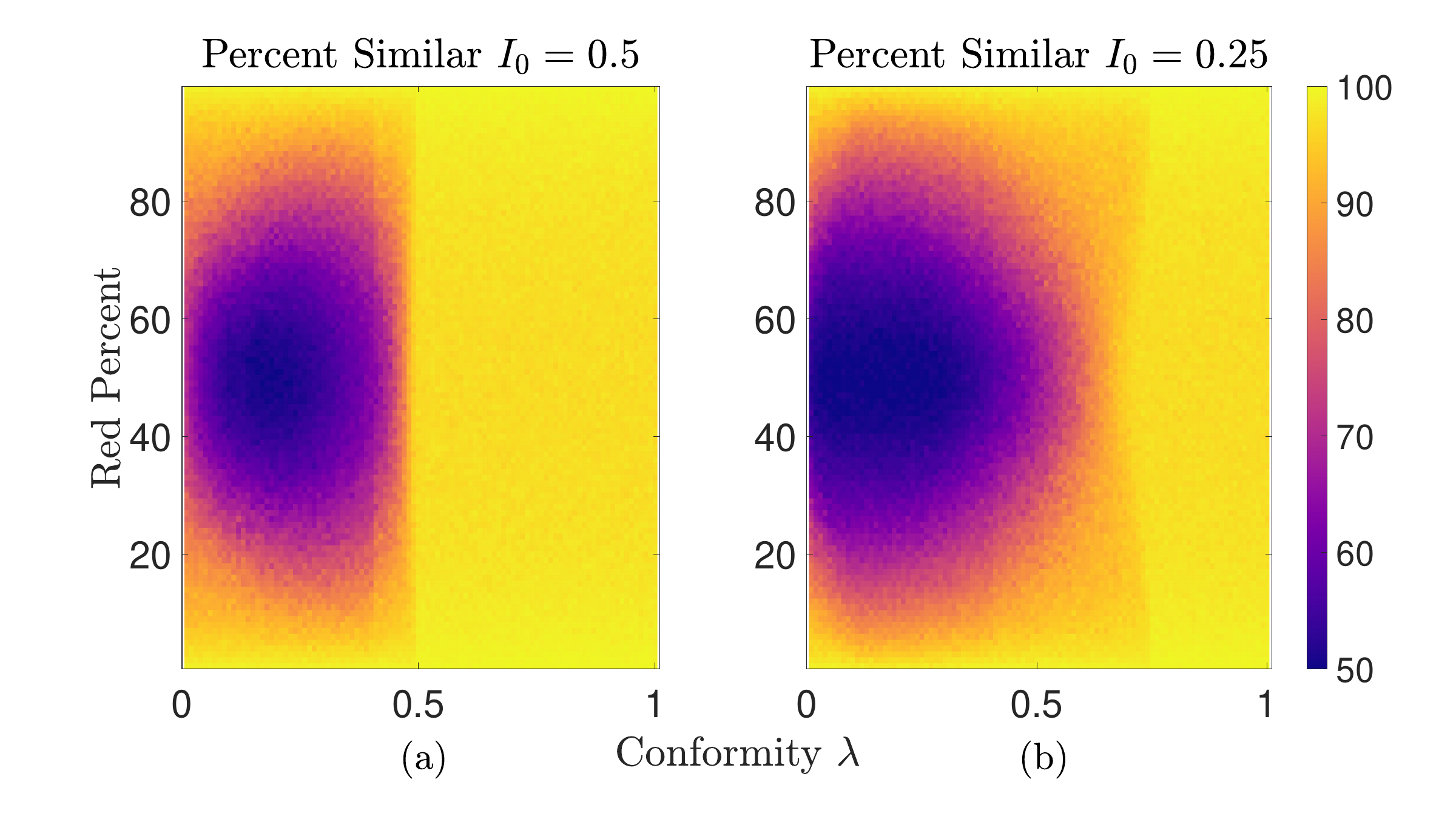}
  \caption{\textbf{The effect of Population Ratio on Segregation.} Equilibrium values of segregation measured by percent similar as the percent of red agents and conformity $\lambda$ varies for initial intolerance values $I^0=25,50$. As the population becomes more imbalanced the amount segregation rises. Furthermore, the threshold into a fully segregated equilibrium is consistent across all population ratios. These thresholds can be matched to the thresholds in panel (a) of \figref{fig:PercentSimandFinIntol}. The parameters for the simulations pictured above are $I^0 = 50$, \textbf{density} = 0.875, and \textbf{non-racial-move} = 0.05.   }
  \label{fig:RedPercent_PerSim}
\end{figure}

The effect of population ratio is illustrated in \figref{fig:RedPercent_PerSim}. An imbalanced population ratio leads to an increase in segregation. This result is not surprising as the even when agents are placed randomly there is a larger chance for an agent from the larger group to have neighbors from the same group. 

As shown in \figref{fig:PercentSimandFinIntol} the threshold for the transition to a completely segregated state when $I^0 = 25, 50$  is $\lambda \approx 0.75$ and $\lambda \approx 0.5$ respectively.  Panels (a) and (b) demonstrates that this conformity threshold is consistent for all population ratios. Therefore while the population ratio affects the degree of integration/segregation for lower conformity levels, the population ratio does not affect where the transition to completely segregated occurs.

\subsection*{Media Influence}

Next, we examined the effect that media has on the development of segregation. In this section, we explore adjusting the media influence values $g(j)$ for both the red and green populations to see the effect that purely positive (tolerant) and purely negative (intolerant) media has on segregation. Additionally, we investigate the effect that heterogeneous media consumption (one population consuming negative media and the other positive, and vice-versa) has on segregation.

We analyze the effect of media influence $g(j)$ on segregation and average intolerance. In the following simulations, here we set the media value to $g(j) = g_R$ for red agents and $g(j) = g_G$ for green agents. We set the initial intolerance to $I^0 = 50$ in across all following runs. The population is split evenly between red and green agents and the \textbf{density} is set to 87.5 percent.

\begin{figure}[H]
    \centering
    \includegraphics[width = \textwidth, height = 75mm]{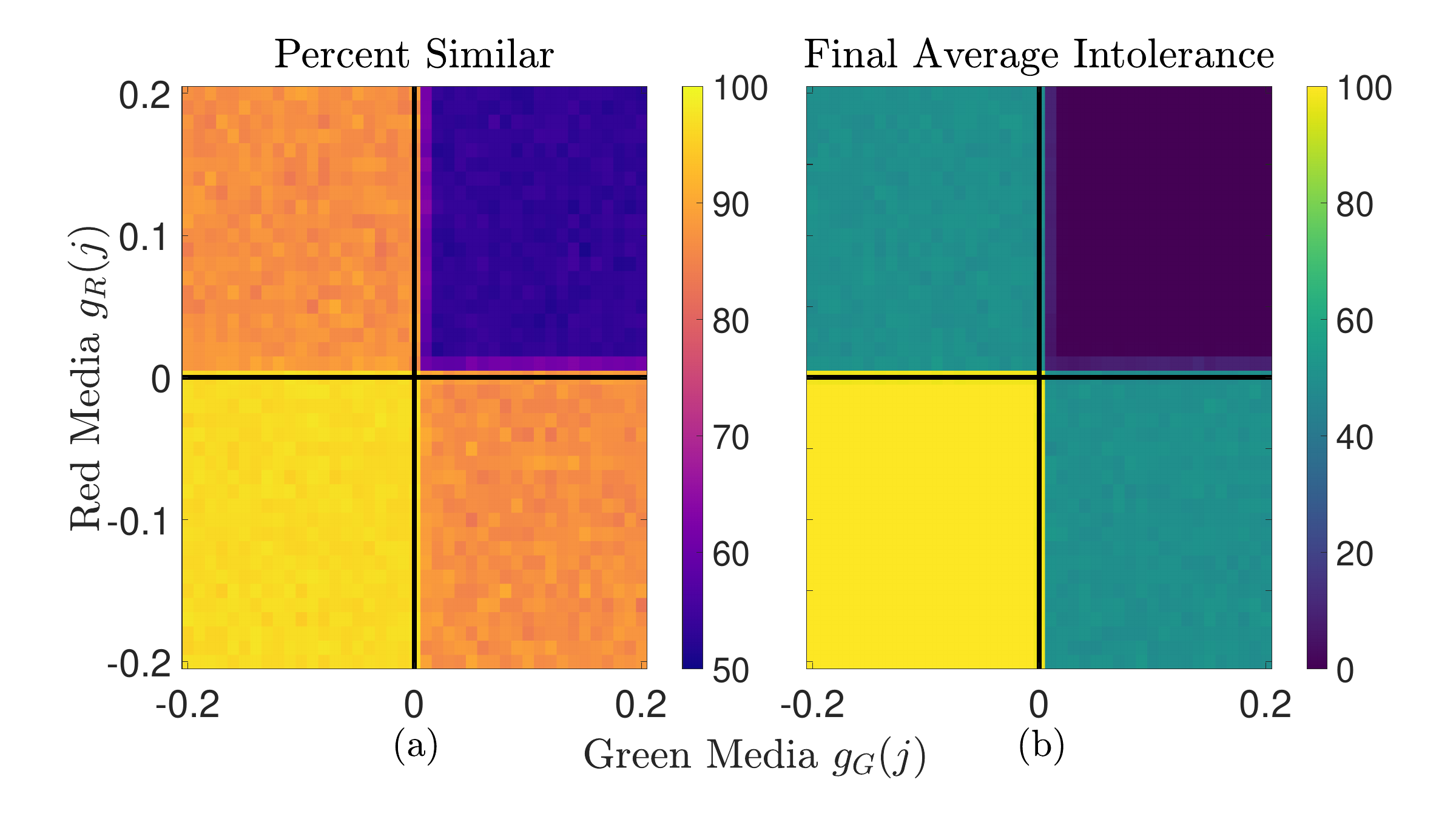}
  \caption{\textbf{Media Effect on Segregation and Intolerance When $\lambda = 0.5$.} Equilibrium values for (a) segregation and (b) final intolerance. When the media values for both the red and green populations consume tolerant media, the integrated equilibrium arises and the intolerance levels approach zero. Here, $I^0 = 50$, \textbf{density} = 0.875, and \textbf{non-racial-move} = 0.05. }
  \label{fig:MediaEffect50width50}
\end{figure}

The effect of media influence on segregation and intolerance is given in \figref{fig:MediaEffect50width50}. Integration develops when at both group's media is positive ($g_G > 0$ and $g_R > 0$).  We can investigate further and check how this result changes if we change the conformity value. We set the conformity value at two values clearly below and above the threshold for initial intolerance $I^0 = 50$ given by \figref{fig:PercentSimandFinIntol}. In the following simulations we set $\lambda = 0.25,0.75$.


\begin{figure}[H]
    \centering
    \includegraphics[width = \textwidth, height = 75mm]{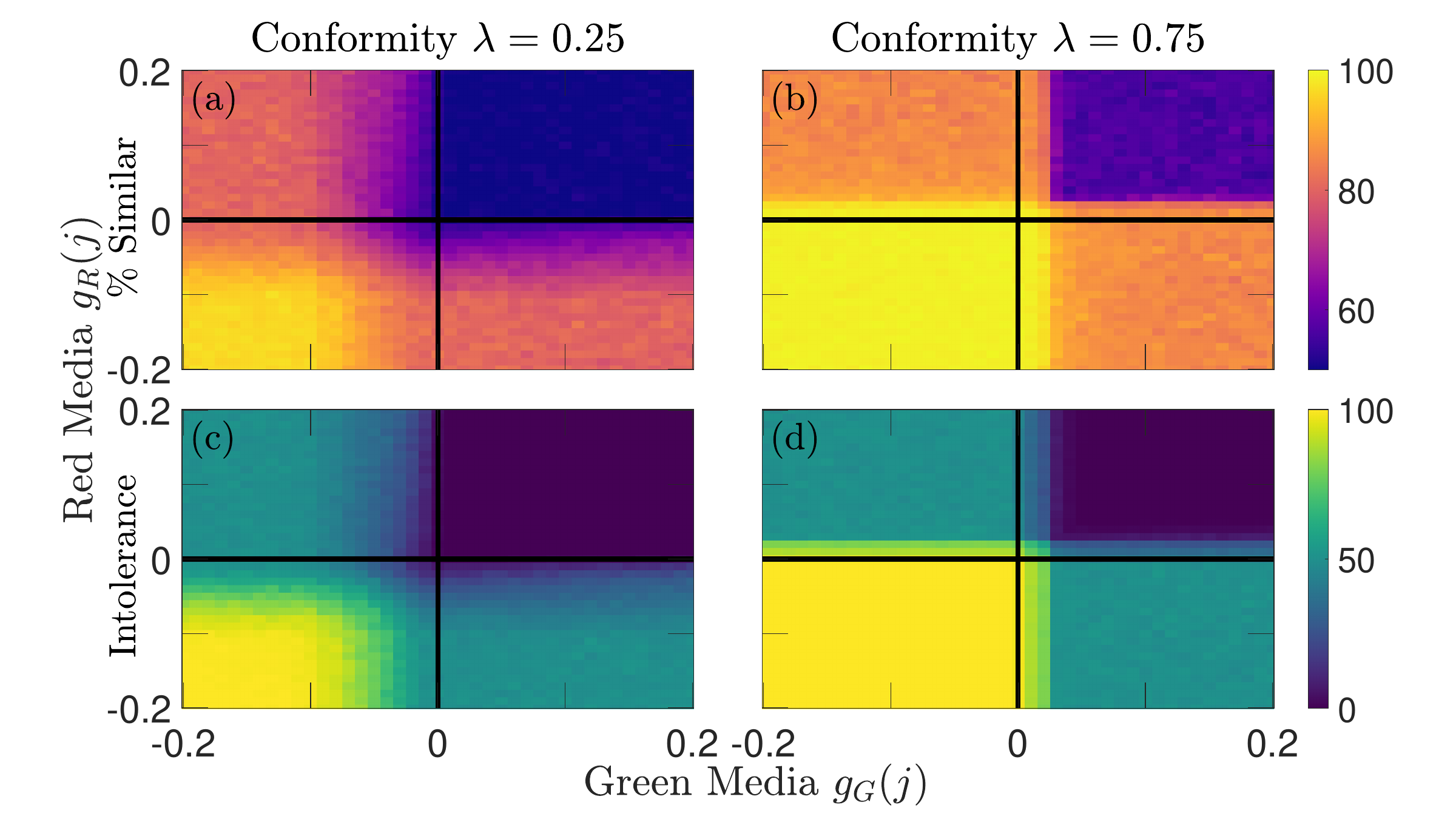}
  \caption{\textbf{Media Effect on Segregation and Intolerance for Conformity Values $\lambda = 0.25,0.75$.} Panel (a) and (c) show the equlibrium segregation and the equlibrium  average tolerance when the conformity is 0.25. As seen in \figref{fig:PercentSimandFinIntol}, when both sets of agents consume neutral media then the equilibrium is an integrated state. This lower value of conformity also allows mitigates the effect that intolerant media has on equilibrium segregation.}
  
  \label{fig:MediaEffect50width75_25}
\end{figure}

The integrated equilibria persists even when both populations consume negative media when $\lambda = 0.25$ as seen in panel (a) of \figref{fig:MediaEffect50width75_25}. Furthermore, transition from an integrated equlibria to a segregated equilibria is far smoother when compared to the case when $\lambda = 0.5$ or $\lambda = 0.75$.  This demonstrates that when agents do not broadly conform to their neighborhood integration can persist in the presence of negative media.
Panel (b) illustrates the case when $\lambda = 0.75$. Here, the segregated state can occur even when both groups are consuming positive media. So, it appears that larger values of conformity reduce the effectiveness of positive media to develop integration.


\subsection*{Non-Racial-Move Probability}

Throughout the previous simulations we set an agent's chance to move for reasons outside of it being unsatisfied with its neighborhood composition to five percent. In this section, we vary the non-racial-move probability to explore the affect moving for reasons outside of neighborhood composition has on segregation and intolerance. In the following simulations, the agents' initial intolerance value are set to $I^0 = 50,25$ and the density is set to 87.5 percent. In the following simulations we vary the non-racial-move probability from 0 to 100. 
\begin{figure}[H]
    \centering
    \includegraphics[width = \textwidth, height = 75mm]{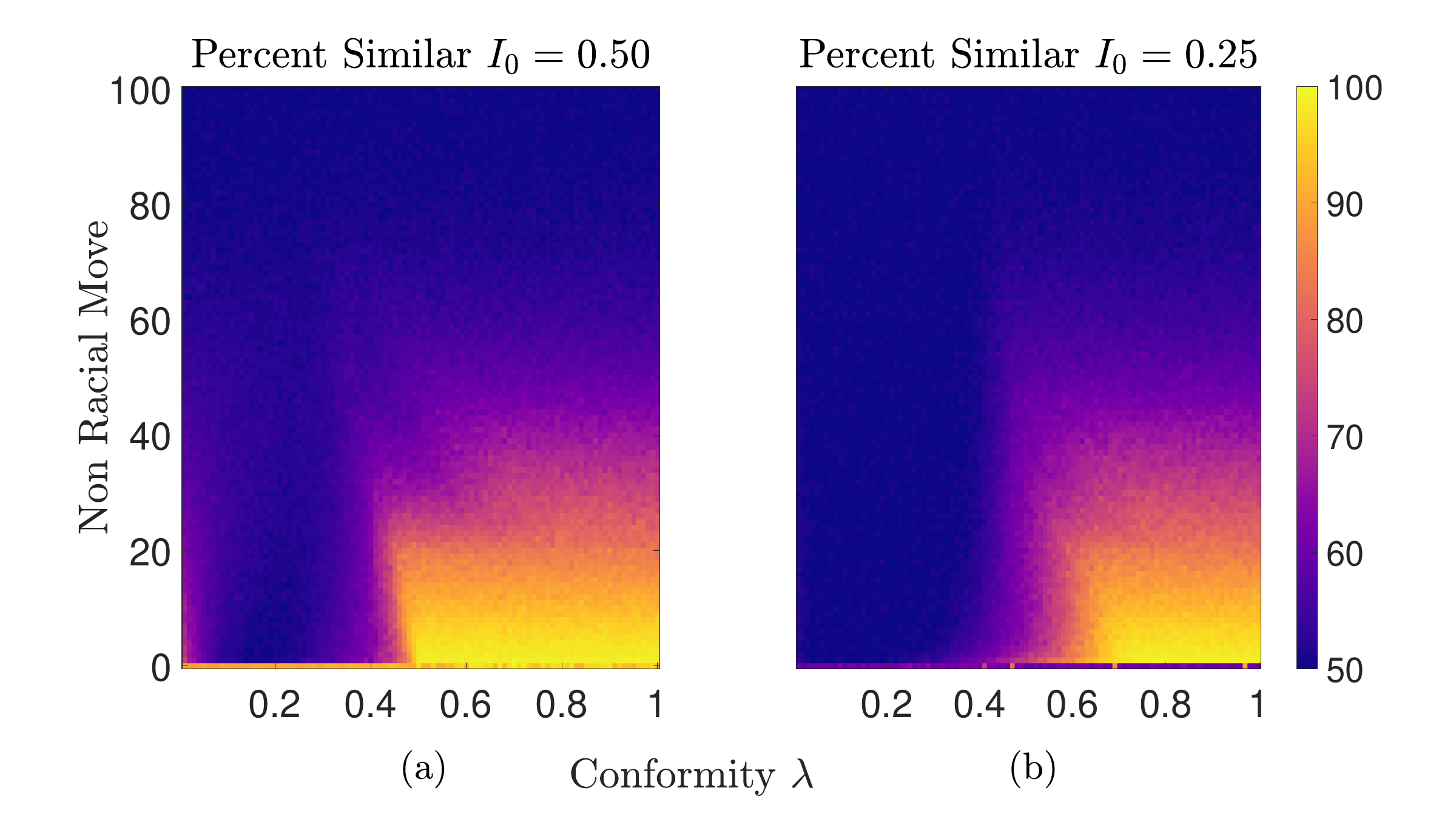}
  \caption{\textbf{The Effect of Moving Due to Non Neighborhood Factors.} Equilibrium segregation and final average intolerance values given the conformity $\lambda$ and probability of moving due to non racial factors when the initial intolerance is (a) $I^0 =50$ and (b) $I^0 =25$. The bottom row gives the equilibria with no random movement. The results are distinct from the rest of the simulations that include randomness.}
  \label{fig:MoveChance}
\end{figure}

As illustrated in \figref{fig:MoveChance}, increasing non-racial-move probability reduces segregation. Panels (a) and (b) demonstrate that the conformity threshold marking the transition from the segregated state to the integrated state demonstrated in $\ref{fig:PercentSimandFinIntol}$ is consistent across all nonzero values of non-racial-move probability. The transition at this threshold becomes less stark as one increases the non-racial-move probability until it is imperceptible.

 Note, the bottom row of the heatmap give the results of simulation when there is no random movement. This implies that introducing randomness into simulation regardless of how small changes the results. There is a negative relationship between non-racial-move probability and segregation, there is not a notable difference in the equilibrium for small values of non-racial-move probability (1 to 10 percent). Therefore, setting the non-racial-move probability to five percent does not have a perceptible effect


\section*{Discussion and Conclusions}
In this paper, we extended Schelling's Model to include sociological factors that could lead to integration, such as intergroup contact, media influence, or movement due to factors outside of neighborhood composition. We demonstrated that integration develops when conformity is low---i.e. when individuals tend not to not to adapt their preferences to their neighborhood population. The conformity threshold is dependent on the society's initial intolerance (e.g. perhaps reflecting the potential effect of previous history on current patterns), however when the agents broadly conform to their neighborhood composition segregated equilibria arise regardless of their initial intolerance. This threshold appears whether the initial state is well-mixed or segregated, although when the initial state is segregated relationship between the critical value of the conformity and initial intolerance appears to be linear.

We found that when both groups consume media that paints the out-group positively segregation is eliminated. When agents broadly conform to their surroundings, segregation can persist even when both groups consume positive media---up to a certain threshold. When agents are broadly repelled by their neighborhood composition, integration seemingly persists even when both groups consume negative media until a critical threshold. We also note that adding in a relatively small amount random movement can lead to integration if conformity is below a critical threshold.

There are several improvements that can be made to this model. This model does not attempt to integrate the conditions laid out by Allport for when intergroup contact leads to reduction in prejudice---equality in status, groups having common goals, and mechanisms assisting in integroup contact---rather these conditions are assumed to already be in place. Further research will be loosening this assumption. It would be of interest to collect data for We leave these alterations to the model for further research. Additionally, adding elements such as physical barriers or spots that do not accept certain groups are next step that one could incorporate into this model. These elements would active steps in segregating races such as parallel bulldozing neighborhoods to place highways to separate ethnic groups and redlining.


Within our framework, it appears that how willing people are to conform is the critical component in the formation of segregation. This might point to a further analysis of the kinds of people who choose to self-segregate and the various methods of altering people's desire to conform so that integration can develop. Furthermore, our work suggests that focusing our supplying positive media of ethnic groups would lead to positive steps in moving toward an integrated populace. We hope that our work leads to more investigation into factors that could drive the population into a more integrated society.


\nolinenumbers

\bibliography{plos}

%
%
%





\end{document}